\documentclass{article}
\usepackage{arxiv}
\usepackage[utf8]{inputenc} 
\usepackage[T1]{fontenc}    
\usepackage{hyperref}       
\usepackage{url}            
\usepackage{booktabs}       
\usepackage{amsfonts}       
\usepackage{nicefrac}       
\usepackage{microtype}      
\usepackage{lipsum}		
\usepackage{graphicx}
\usepackage{doi}
\usepackage{cite}
\usepackage{amsmath,amssymb,amsfonts}
\usepackage{algorithmic}
\usepackage{amsmath}
\usepackage{textcomp}
\usepackage{multirow}
\usepackage{wrapfig}
\usepackage{tabularx}
\usepackage{ragged2e}
\usepackage{colortbl}
\usepackage{xcolor}
\usepackage{hyperref}

\usepackage{pifont}
\usepackage[super]{nth}
\usepackage{soul}
\usepackage[numbers]{natbib}

\usepackage{filecontents} 

\title{Dynamic Resource Partitioning for Multi-Tenant Systolic Array Based DNN Accelerator}

\author
 { 
{
Midia Reshadi}\thanks{This paper was accepted at $31^{st}$ Euromicro International Conference on Parallel, Distributed, and Network-Based Processing (PDP 2023).} \\
	School of Computer Science and Statistics\\
	Lero, Trinity College Dublin\\
	Dublin 2, Ireland \\
	\texttt{Midia.Reshadi@tcd.ie} \\
	\And
	{
	\hspace{1mm}David.Gregg} \\
	School of Computer Science and Statistics\\
	Lero, Trinity College Dublin\\
	Dublin 2, Ireland \\
	\texttt{David.Gregg@tcd.ie} \\
}

\hypersetup{
pdftitle={A template for the arxiv style},
pdfsubject={q-bio.NC, q-bio.QM},
pdfauthor={David S.~Hippocampus, Elias D.~Striatum},
pdfkeywords={First keyword, Second keyword, More},
}

\begin{document}
\maketitle

\begin{abstract}
Deep neural networks (DNN) have become significant applications in both cloud-server and edge devices. Meanwhile, the growing number of DNNs on those platforms raises the need to execute multiple DNNs on the same device.
This paper proposes a \textit{dynamic partitioning algorithm} to perform concurrent processing of multiple DNNs on a systolic-array-based accelerator. Sharing an accelerator's storage and processing resources across multiple DNNs increases resource utilization and reduces computation time and energy consumption. 
To this end, we propose a \textit{partitioned weight stationary} dataflow with a minor modification in the logic of the processing element.
We evaluate the energy consumption and computation time with both heavy and light workloads. Simulation results show a 35\% and 62\% improvement in energy consumption and 56\% and 44\% in computation time under heavy and light workloads, respectively, compared with single tenancy.
\end{abstract}


\section{Introduction}
{\hskip 1em}Deep neural networks have permeated applications like recommender systems \cite{karatzoglou2017deep}, self-driving cars \cite{chowdhuri2019multinet}, and language translation \cite{lewis2019bart}. The massive demand for DNN processing has led designers to develop domain-specific hardware accelerators to achieve energy efficiency and sufficient processing capacity \cite{chen2016eyeriss}\cite{kwon2018maeri}\cite{du2015shidiannao}. Applications like augmented and virtual reality (AR/VR) \cite{wu2019machine} and autonomous driving cars use several DNNs for different sub-tasks in edge-side devices. For example, a VR application includes hand pose estimation and hand and eye-tracking sub-tasks as user inputs \cite{Rabii2019}.
Similarly, cloud infrastructure \cite{fowers2018configurable} offers INFerence-as-a-Service (INFaaS) \cite{romero2019infaas} for processing multiple real-life DNN-based applications in parallel. Thus, multi-tenant accelerators are shared across multiple requests on cloud-computing platforms.

There are two different methods of running multiple DNNs on the accelerator: sequential and concurrent. In the sequential approach, a layer from one DNN executes on the accelerator at any time. In the concurrent approach, multiple layers from different DNNs can be executed in parallel on the DNN accelerator.
The goal of this paper is to enable parallel processing of more than one DNN layer on the systolic array accelerator, where we share storage and computing resources across multiple DNNs.

\begin{figure}[]
\centering
\centerline{\includegraphics[scale=0.6]{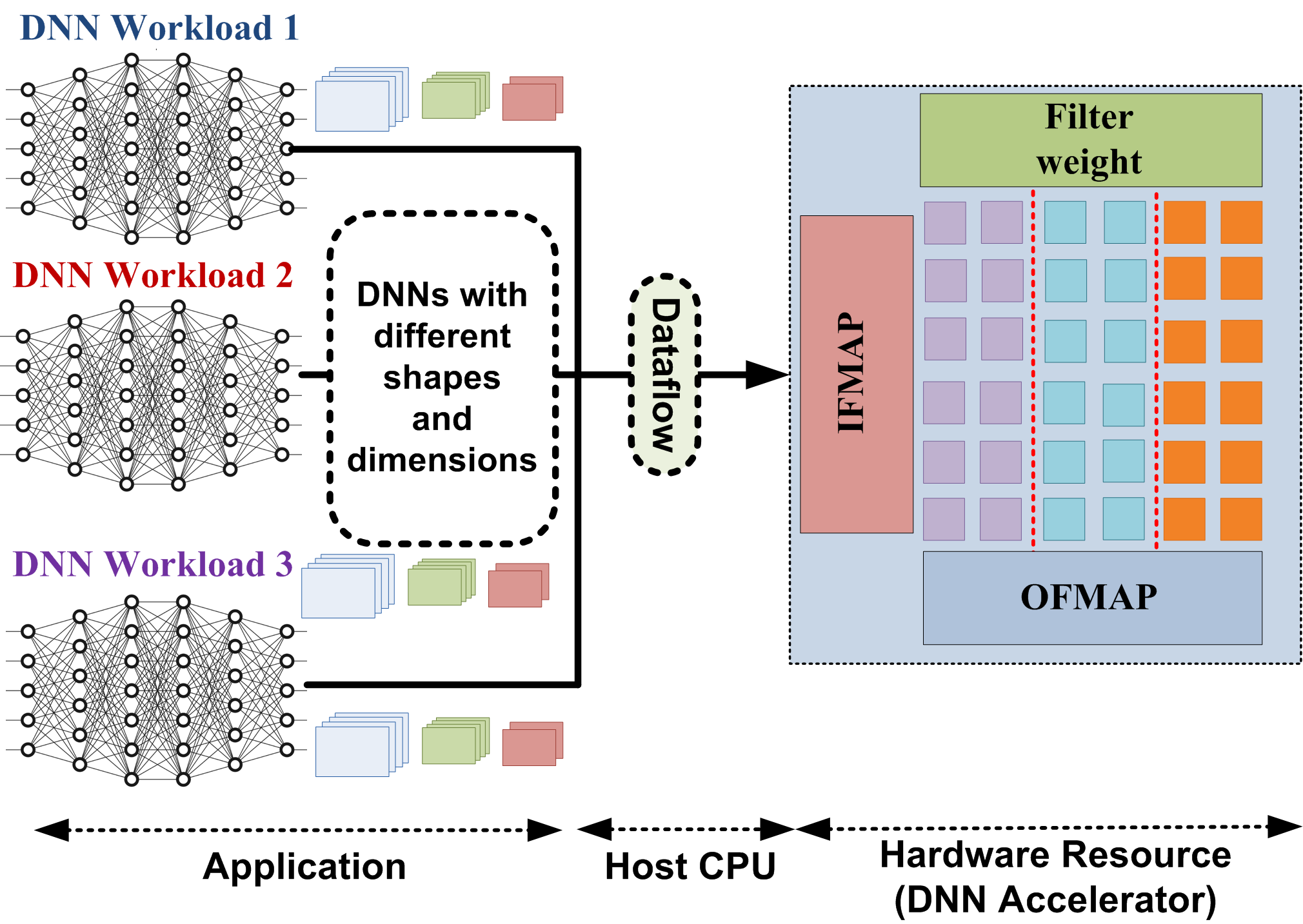}}
\caption{The multi-tenant DNN accelerator.}
\label{fig.1}
\end{figure}

In processing each DNN layer, the dataflow procedure defines scheduling, bounding, and assignments of tensor values to memory and computation resources \cite{chen2017using}. 
The main goal of the dataflow procedure is to maximize data reuse to achieve energy efficiency. 
Three basic dataflow approaches are commonly proposed for implementing DNNs on systolic-arrays: \textit{weight} stationary, \textit{input} stationary, and \textit{output} stationary. The weight stationary approach pre-loads weight data to each of the processing elements (PEs) of the systolic array, and then streams inputs data through the systolic array to produce a stream of outputs. The input-stationary approach is similar to weight-stationary, but the role of weights and inputs is swapped. In the output stationary approach, inputs and weights are streamed through the systolic array to compute output values at each of the PEs, after which the outputs must be drained in a separate stage.

Fig. \ref{fig.1} shows an example of three DNNs that are mapped to a weight-stationary systolic array. The role of the dataflow
procedure is to allocate work to the hardware resources, and map weight, input, and result data to the PEs.
First, the weight values are moved from the filter weight buffer into the individual processing elements of the systolic array. In the next stage, input data is streamed from the input feature map (IFMAP) buffer from left to right across the rows of the systolic array. Each input value is multiplied by the weight value stored within the corresponding PE, and the product flows downward to the PE vertically below. As the products flow downward vertically through the PEs, they are added together, so that each PE produces a partial sum of products each cycle. The partial sums move downward vertically along the columns of the PEs, until the total sum emerges at the bottom of the systolic array, and is stored to the output feature map (OFMAP) buffer.

This paper proposes a dynamic partitioning algorithm as a solution at the data-mapping level to improve resource sharing across multiple DNN layers.
This algorithm needs a small change in PE logic, but it requires no expensive hardware costs, such as clusters with additional inter-cluster communication. Resource partitioning means allocating parts of the data to subsets of resources (processing and storage components). This operation is inherently data-level management and can be implemented in the dataflow procedure.
We partition the systolic array vertically, so that partial sums from the same layer are added together. There is no easy way to partition our weight-stationary systolic array horizontally, because partial sums always flow downward towards the OFMAP, and partial sums from different layers must be kept separate.

Resource partitioning increases resource utilization and improves energy efficiency and computation time. We summarize the main contributions of this paper as follows:
\begin{figure}[]
\centerline{\includegraphics[scale=0.44]{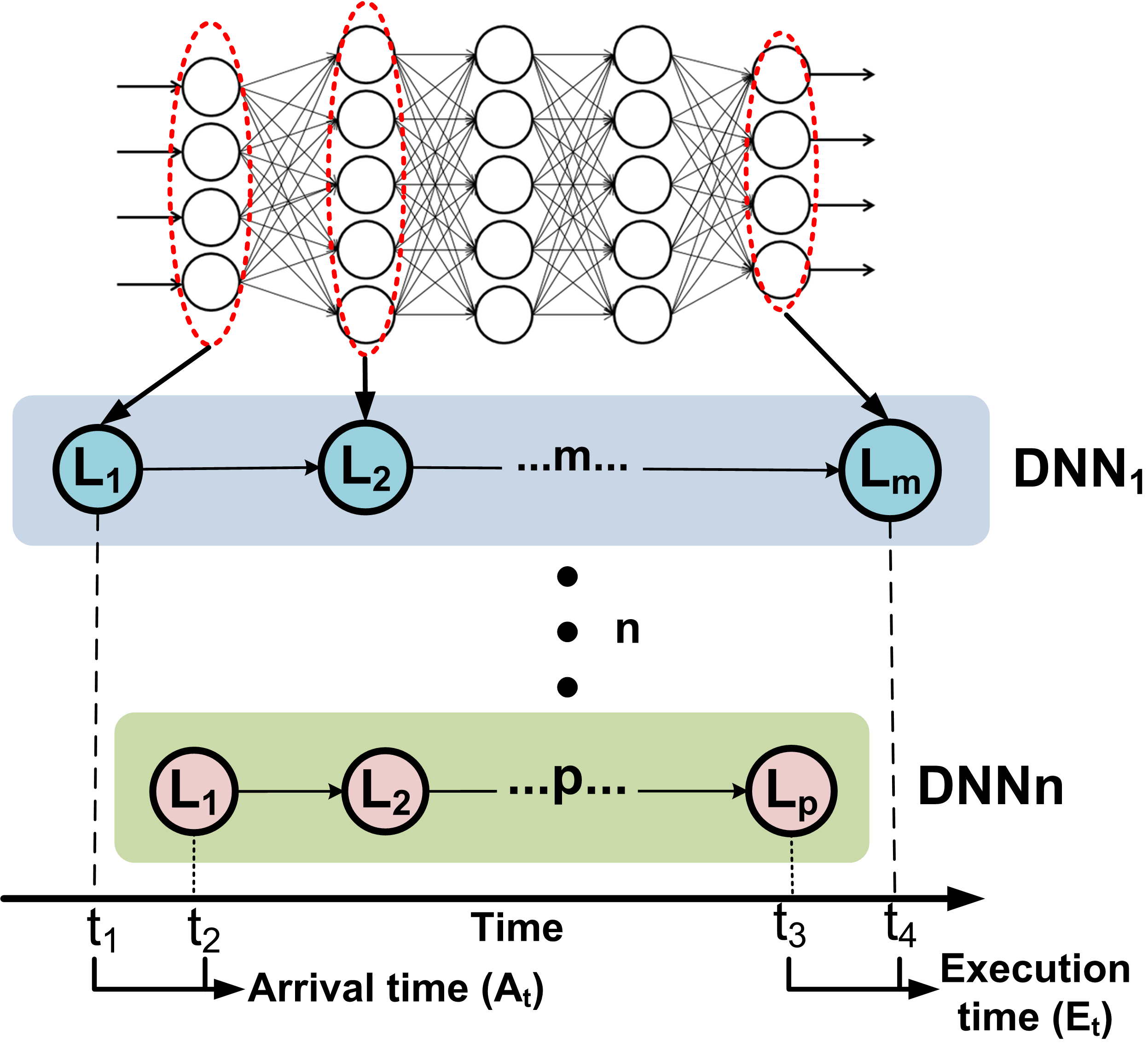}}
\caption{The DNNG graph consisting of $n$ DNNs.}
\label{fig.2}
\end{figure}
\begin{itemize}
\item \textbf {Dynamic resource partitioning of multi-tenant DNN accelerator.} This paper applies dynamic resource partitioning to the PE and memory elements to increase resource utilization in multi-tenant or multi-DNN acceleration platforms. 
\item \textbf {Partitioned weight stationary mechanism.} This paper extends the weight stationary mechanism for supporting dynamic resource partitioning.
\item \textbf {Data mapping solution with a slight hardware modification.} We simply add a tri-state gate to the processing element logic to provide the calculation control required to enable the \textit{partitioned weight stationary} dataflow.
\end{itemize}

This paper is organized as follows: Section II provides preliminaries on DNN accelerators; Section III presents the dynamic partitioning algorithm in detail; Section IV presents comprehensive evaluations; Section V concludes.

\section {Preliminaries}
{\hskip 1em}The primary purpose of this section is to define key concepts in a multi-tenant deep neural-network accelerator. We introduce the structures of the multi-DNN workloads and systolic array logic in detail.
\subsection{Deep Neural Networks Graph (DNNG)}
{\hskip 1em}Multi-DNN workloads comprise several deep neural networks (DNN), including various layers with different dimensions. Accordingly, we introduce the \textit{deep neural network graph (DNNG)} for the abstract definition of multi-DNN workloads. A deep neural network graph is a weighted directed acyclic graph (DAG), \(G(V,E)\) where each vertex \(v_i\in V\) corresponds to layer \(l_i\in L\) that \(L=\)\{$l_1$,$l_2$,\ldots,$l_m$\}. The edge $e_i\in E$ corresponds to the flow of data between layers, and therefore defines the precedence of layer execution. Each DNNG has an \textit{arrival time} $A_t$ and estimated \textit{execution time} $E_t$. Multi-DNN systems include $n$ DNNs, and each DNN comprises a different number of layers, as Fig. \ref{fig.2} shows the pool of $n$ DNNs with $m$ and $p$ layers.

Each layer consists of three convolution tensors: \textit{filter weights (FW)}, \textit{input feature map (IFMap)}, and \textit{output feature map (OFMap)}, in which each tensor is a 4-dimension array with multiple shapes: $\textbf{FW}\in\mathbb{R}^{MCRS}$, $\textbf{IFMap}\in\mathbb{R}^{NCHW}$, and $\textbf{OFMap}\in\mathbb{R}^{NMPQ}$. Thus, the shapes of each layer are:
\begin{equation}
shapes(l_i)=\{M,N,C,R,S,H,W,P,Q\}
\end{equation}

The number of multiply and accumulate (MAC) operations required to process a layer equals the product of the shapes of FW and IFMaps:
\begin{equation}
Opr(l_i)=M\times N\times C\times R\times S\times H\times W
\end{equation}

Calculating the number of MAC operations is used as a measure to estimate the execution time of each layer.

\begin{figure}[] 
\centering
\centerline{\includegraphics[scale=0.72]{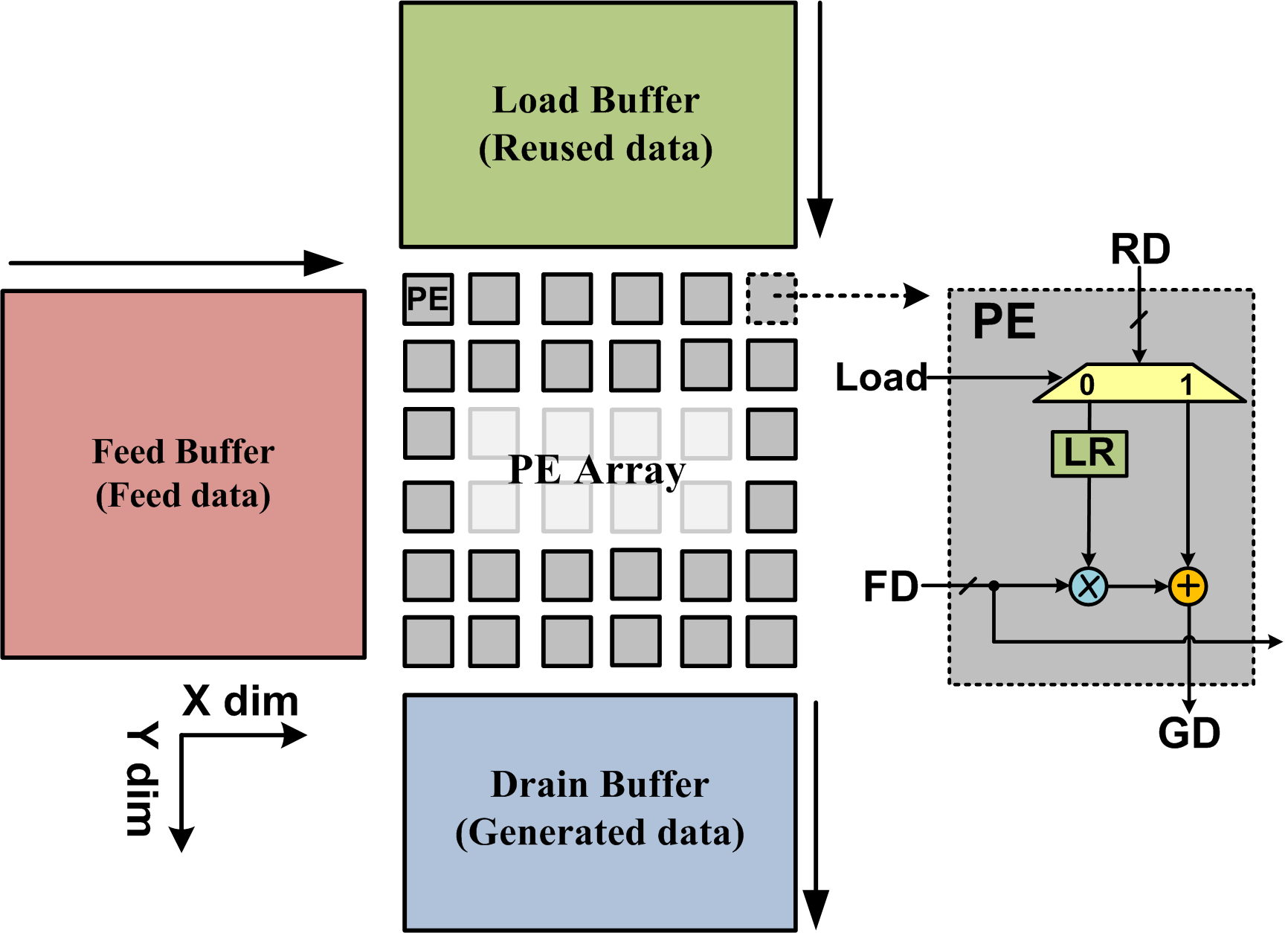}}
\caption{The systolic-array based DNN accelerator. In the PE logic, $LR$, $FD$, $RD$, and $GD$ stand for load register, feed data, reused data, and generated data, respectively. $FD$ and $RD$ are input data ports, and $GD$ is an output data port and $load$ is input control port. $X dim$ and $Y dim$ refer to $X$ and $Y$ dimensions, respectively.}
\label{fig.3}
\end{figure} 

\subsection{Systolic-array Based DNN Accelerator}
{\hskip 1em}A systolic-array-based accelerator (SA) is a domain-specific processor that consists of a 2D array of processing elements (PE) and storage components for matrix multiplication in applications like deep neural network processing.

Our design is similar to the Google TPU \cite{jouppi2017datacenter}, a weight-stationary systolic array with three SRAM memory buffers: one for each of the filter weights, input feature map and output feature map.
However, we use more abstract naming for buffers, that is the \textit{load}, \textit{feed}, and \textit{drain} buffers, as shown in Fig. \ref{fig.3}. For ease of presentation, we represent the dimension of each buffer as $Buffer [row, column]$, which signifies the buffer's row and column.

Each PE comprises a \textit{load register} (LR) and MAC unit to perform multiply and accumulate operations. The PE logic has two operating modes depending on the $load$ input: Load ($load=0$) and Calculate ($load=1$). Fig. \ref{fig.3} depicts the structure of the PE with two data input-ports including \textit{reused-data (RD)}, \textit{feed-data (FD)} and \textit{generated-data (GD)} as a data output-port. There is also a control input-port called $load$ that specify the PE operation. Each PE element is represented as \(PE [x,y]\) where $x$ and $y$ refer to the PE's index in the $x$ and $y$ dimensions, respectively. For instance, \(PE [0,2]\) denotes PE in the first row and third column of the PE array.

In the systolic array, tensor values move through PE-array in $X$ the dimension from the feed buffer and $Y$ dimension from the load buffer. After completing each MAC operations in the PE, the results move downwards to the adjacent PE in the $Y$ dimension. Note that data from the load buffer and results from MAC operations both move downwards in the $Y$ dimension along the same inter-PE connections. Therefore, we must load data from the load buffer and compute results in separate steps. All three on-chip buffers are connected to off-chip DRAM memory.

\section{Dynamic Resource Partitioning Algorithm}
\subsection{Systolic Array Partitioning}
{\hskip 1em}As shown in Fig. \ref{fig.3}, our systolic array comprises computational and storage elements. Resource partitioning means dividing storage and processing elements into equal-sized partitions and employing them to process multiple DNN layers simultaneously. 
This goal is achieved by assigning tensors of a layers to a resource partition. For example, allocating a part of each storage element to tensors along with utilizing a subset of PEs to compute. Fig. \ref{fig.6}(a) shows a simple example of a dividing a systolic array in two partitions. Two DNNs are in task queue called $DNN_1$ and $DNN _2$ with $m$ and $n$ layers, respectively. Two memory spaces of $load$, $feed$, and $drain$ buffers are allocated to the DNN layers. Furthermore, the PE-array is split into two $6\times 3$ sub-arrays (six rows and three columns).
\begin{figure}[]  
\centering
\centerline{\includegraphics[scale=0.6]{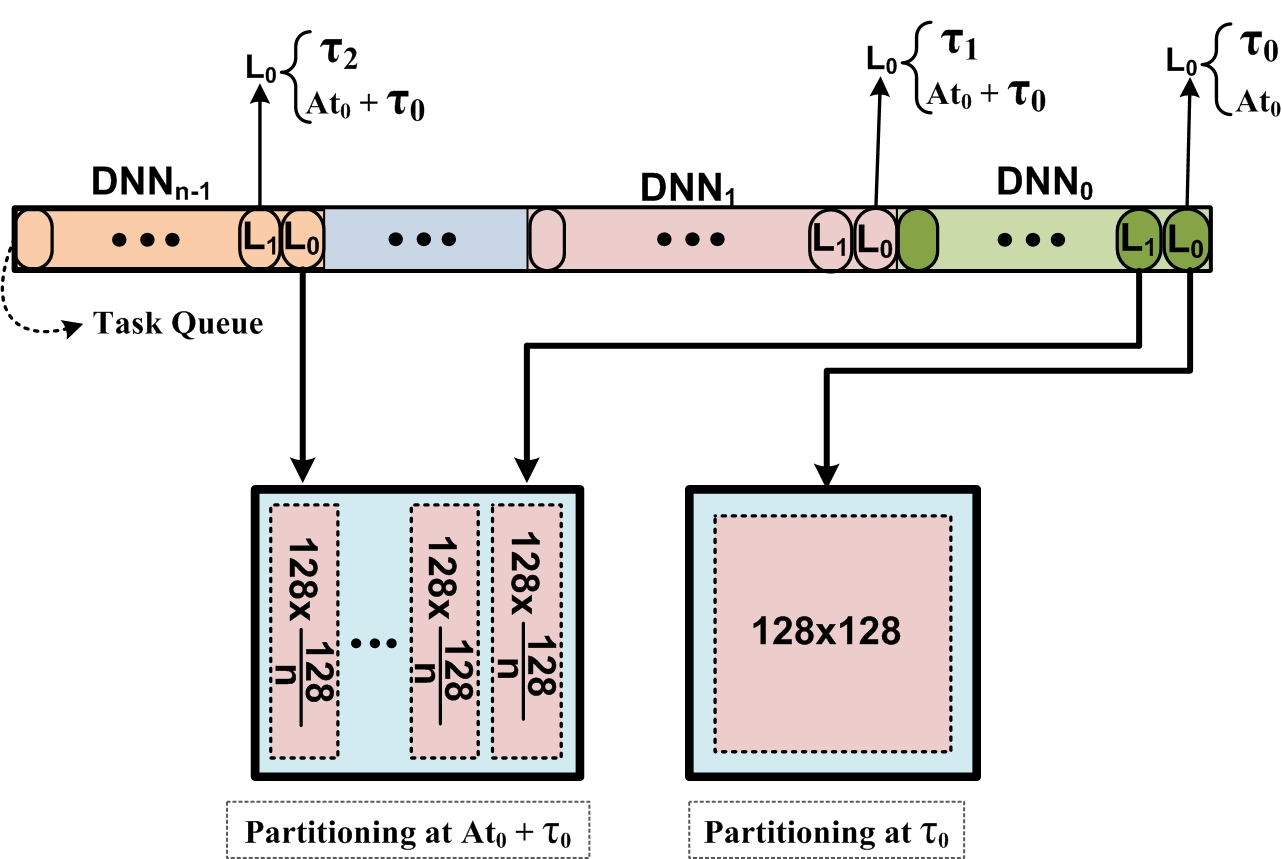}}
\caption{The partitioning mechanism. $A_t$ is arrival time and $\tau_0, \tau_1, \tau_2$ are execution time ($E_t$)}
\label{fig.4}
\end{figure}     
\subsection{Partitioning Strategy}
{\hskip 1em}The systolic array structure shown in Fig. \ref{fig.3} allows only vertical partitioning. For example, in a $128\times128$ systolic array ($128$ PE rows and $128$ PE columns), if the number of partitions is 4, then the partition shapes will be $128\times32$ ($128\times\frac{128}{4}$). Using horizontal partitioning (e.g., two $64\times128$ partitions) causes incorrect results because the OFMap results of the upper partition enter the lower partition as inputs. Fig. \ref{fig.4} shows the $n$ number of DNNs as available tasks for processing in a task queue. The first arrival time belongs to the first layer of the first DNN ($L_0$ of $DNN_0$); the partitioning algorithm assigns the first layer to the entire systolic array, so, the first layer is processed by all PEs because there are no other available layers to process in parallel with this layer. 

As it is shown in Fig. \ref{fig.4}, after finishing the layer $L_0$ of $DNN_0$ at time $At_0+\tau_0$, there are $n$ available layers in the queue for processing ($L_1$ of $DNN_0$, $L_0$ of $DNN_1$,.., $L_0$ of $DNN_{n-1}$).

Therefore, the accelerator is partitioned into $n$ parts with the PE-array size equal to $128\times \left \lfloor \frac{128}{n} \right \rfloor$ (PE row=$128$, PE columns=$\left \lfloor \frac{128}{n} \right \rfloor$). Some DNNs consist of only a few layers and are finished sooner than the DNNs with many layers. Hence, some partitions are freed after completing its allocated layers, and then these partitions may be merged if they are adjacent. Partition merging accelerates the processing of the remaining layers with more processing resources.
\begin{figure}[] 
\centering
\centerline{\includegraphics[scale=0.2]{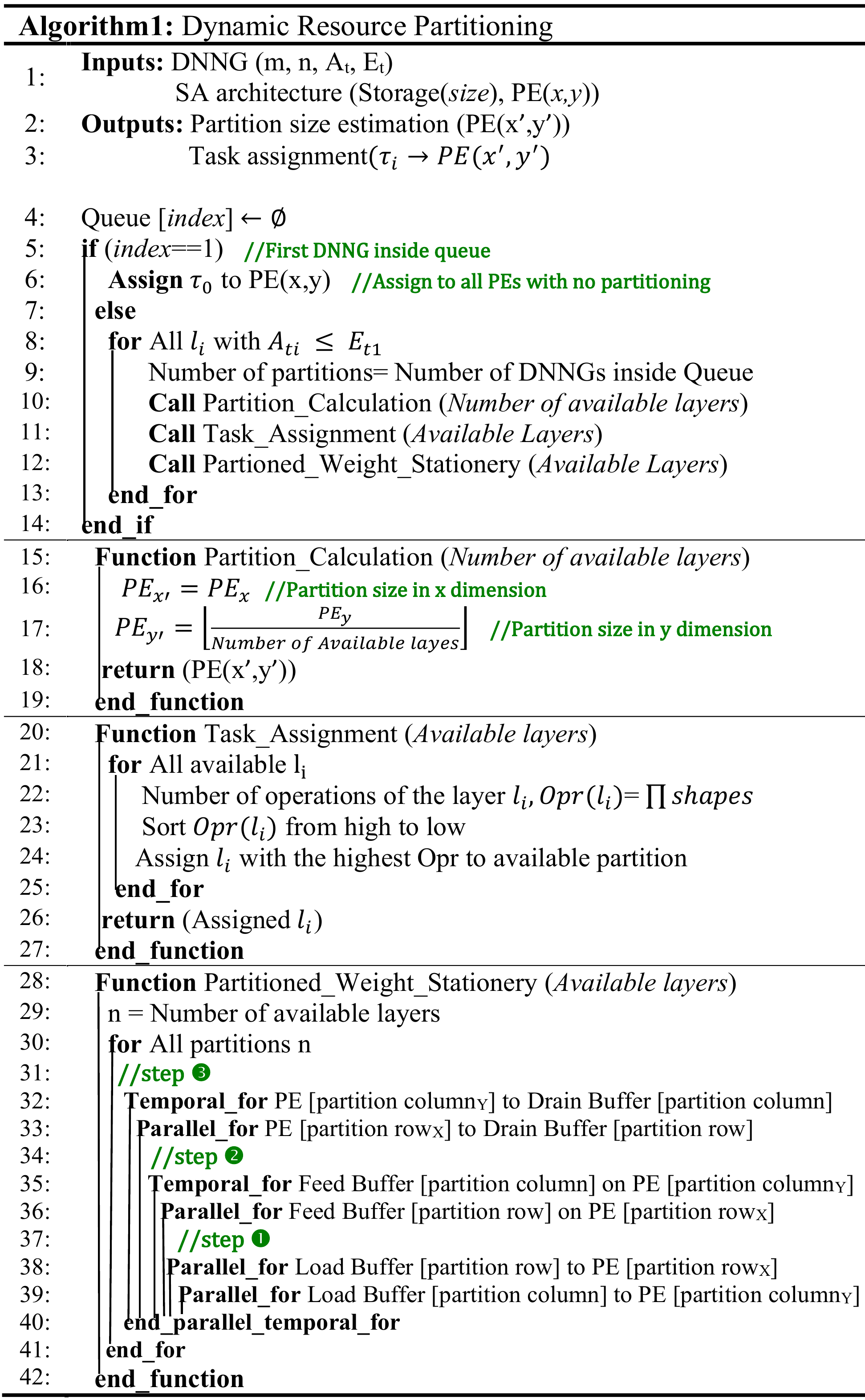}}
\caption{Dynamic resource partitioning algorithm}
\label{fig.5}
\end{figure}    
\subsection{Partitioning Algorithm}
{\hskip 1em}Fig. \ref{fig.5} shows the pseudo-code of the dynamic partitioning algorithm. The algorithm takes the DNN graph (DNNG) and the specs of the accelerator as inputs (Line:1) and, after estimating the partition size (Line:2), assigns layers to the partitions operated as sub-accelerators (Line:3).

All DNNs are stored in the queue, and the first layer of the first DNN is run by all PEs (Line:6). More DNNs join the queue after processing the first layer. Based on waiting layers in the queue whose arrival time is lower than the execution time of the first layer (Line:8), the \emph{Partition\_Calculation} function (Line:15-19) computes the size of partitions, and the \emph{Task\_Assignment} function (Line:20-27) assigns the available layers to the partitions.

The \emph{Partition\_Calculation} function divides PEs only in the $Y$ dimension; the height of the partition is always equal to the height of the base PE-array, but the width of PE-array is divided by the number of available tasks (Line:17).
\begin{figure}[]
\centering
\centerline{\includegraphics[scale=0.72]{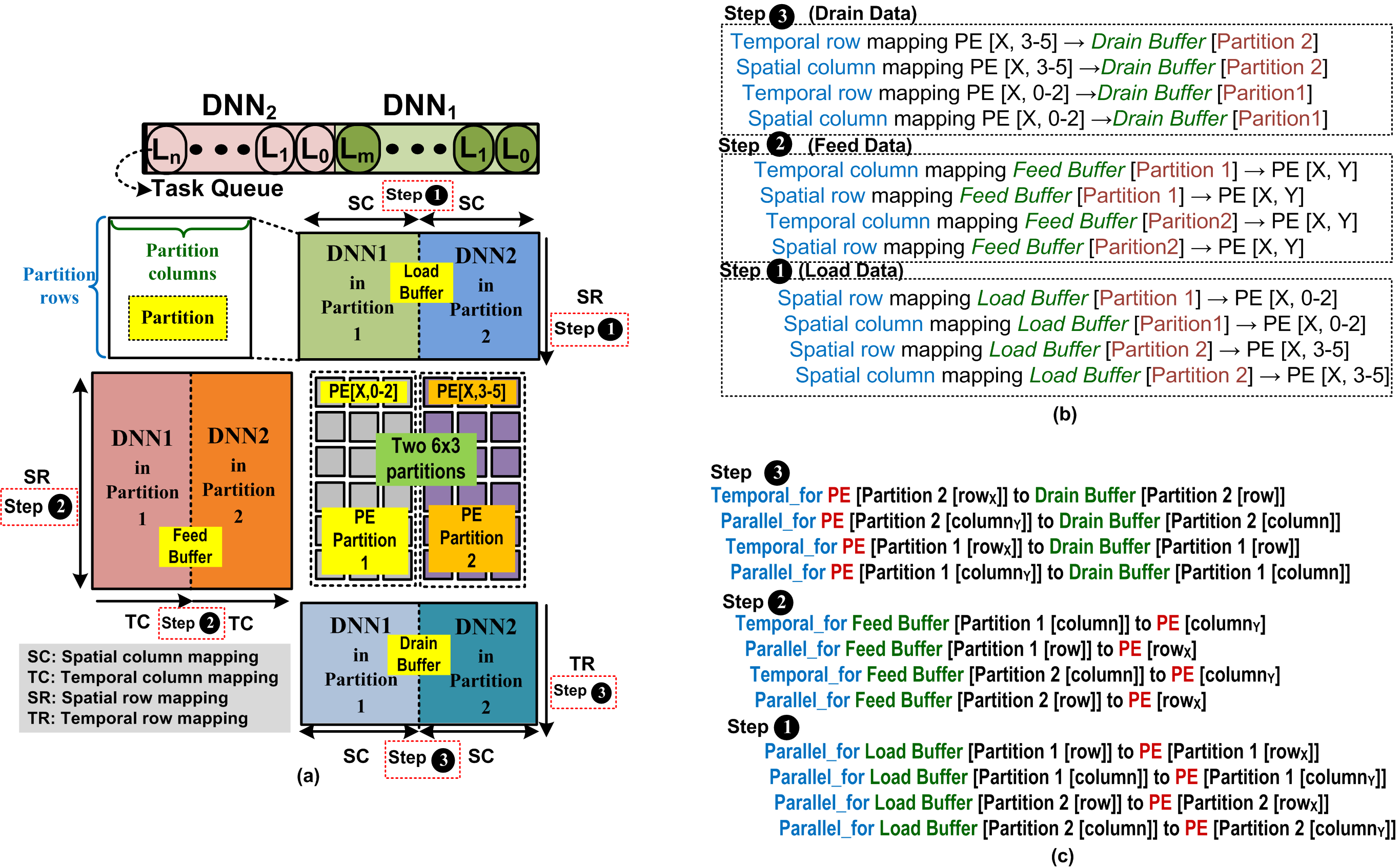}}
\caption{(a) Partitioned systolic array. In each PE partition, $PE[X, 0-2]$ means PE address in the entire $X$ range ($X=0\ to\ 5$) and $Y$ in the range of $0$ to $2$. (b) Symbol-based representation of partitioned weight stationary dataflow (c) Loop-nest definition of of partitioned weight stationary dataflow.}
\label{fig.6}
\end{figure}  

In the \emph{Task\_Assignment} function, all available layers are sorted based on the required MAC operations (Eq. (2)) to prioritize layers with higher processing demand (Line 22-23). Prioritizing layers is necessary because, in some cases, we may have partitions with different sizes due to merging some partitions. At the same time, there may be several layers with different dimensions ready for processing. In this case, layers with higher dimensions are assigned to the partitions with higher resources. Finally, the partitioned weight stationary function (Line:28-42) is called to perform dataflow operations on the partitioned resources.
\begin{figure}[]  
\centering
\centerline{\includegraphics[scale=0.4]{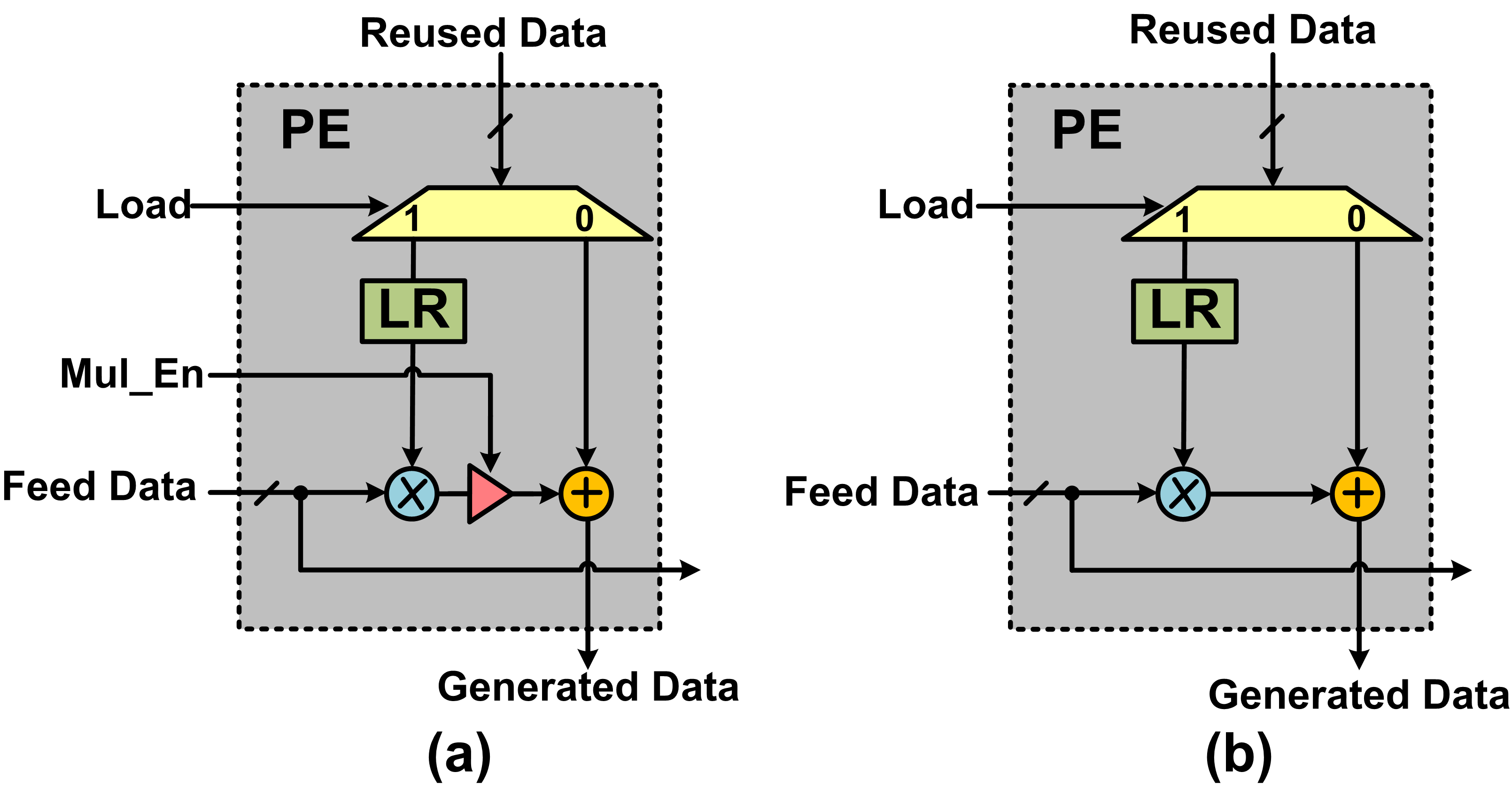}}
\caption{(a) The proposed PE with a controlled multiplier. $Mul\_En$ is connected to the $En$ (enable) signal of the tri-state gate. In the state of $Mul\_En=0$, the multiplier is disconnected from the adder logic, so the PE just passes data in the $X$ dimension. When $Mul\_En=1$, PE operates in conventional (calculate) mode. (b) The Baseline PE.}
\label{fig.7}
\end{figure}     

\subsection{Partitioned Weight Stationary Dataflow}
{\hskip 1em}Dataflow defines the temporal and spatial scheduling of operations for computation \cite{chen2017using}. Simply put, operation scheduling is carried out in space and time.
In this work, we propose a \textit{partitioned weight stationary} dataflow to support the partitioning algorithm. We label the three steps of the dataflow with abstract names that are analogous to internal buffers: \ding{182} \textit{load}, \ding{183} \textit{feed}, and \ding{184} \textit{drain} steps. As it is shown in Fig. \ref{fig.5} (Line: 28-42), all steps are defined by the loop-nests.

\textbf{Data Load.} In step \ding{182}, the reused-data (filter weights) partitions are assigned to PE partitions from the load buffer. 
Hence, the $load$ input of all PEs inside a partition is set to $1$ ($load=1$) to store the reused-data in the load register (LR).
Step \ding{182} includes spatially assigning the rows and columns of reused data (that is \textit{stationary}) from the load buffer to the $X$ and $Y$ dimensions of PE partitions. Hence, the load operation is defined by two spatial maps as shown in Fig. \ref{fig.5} (Line: 38-39). 

Typically, \texttt{\textbf{Parallel\_for}} \cite{reshadi2021local} denotes the spatial mapping of data in a loop-nest representation, which means assigning tensor values to the set of PEs. In contrast, \texttt{\textbf{Temporal\_for}} denotes temporal assignment, which means assigning tensor values move across the systolic array cycle by cycle \cite{kwon2019understanding}. 

In Fig. \ref{fig.6}(a) step \ding{182} involves spatial row and column assignments of reused-data to PE partitions, and Fig. \ref{fig.6}(b) defines all assignments with details of PE-array indices during data assignments.
Fig. \ref{fig.6}(c) shows a for-loop definition, including two \texttt{\textbf{Parallel\_for}} for each partition assignment.

\textbf{Data Feed.} In step \ding{183}, the partitions of feed data (IFMap) move from feed buffers across the PE arrays in the $X$ dimension, and PEs calculate partial sum and then send them to the neighbouring PEs in the $Y$ dimension. 
As a result, the data movement is performed by spatially assigning feed data to PE arrays in rows and temporally assigning feed data to PE arrays in columns. Fig. \ref{fig.5} (Line:35-36) shows an abstract for-loop presentation of step \ding{183}.
In contrast to step \ding{182}, feed data partitions pass through multiple PE partitions. Thus, a technique is required to guarantee that each feed data partition is computed solely by its matching PE partition. To create such a feature with minimum hardware overhead, we just add a tri-state buffer between adder and multiplier in PE logic. Fig. \ref{fig.7}(a) shows our proposed PE with controlled-multiplication ability compared to the baseline PE (Fig. \ref{fig.7}(b)). The proposed PE has a second control input called $Mul\_En$ which is used to choose to enable or disable multiplication when passing through relevant or irrelevant partitions.
Therefore, in step \ding{183} the $load$ input of all PEs is set to $0$ ($load=0$) to operate in calculation mode and the $Mul\_En$ input of PEs is set to 1 ($Mul\_En=1$) only when they pass trough the corresponding data partition. For example, in Fig. \ref{fig.6}(a) the $Mul\_En$ input of partition 1 PEs is set to 0 ($Mul\_En=0$) when passing DNN2 data in step \ding{183}, and it is set to 1 ($Mul\_En=1$) in partition 2 PEs when passing DNN2.

\textbf{Data Drain.} Finally, in step \ding{184}, the generated OFMaps (drain data) leave PEs by spatial column and temporal row assignment, which is shown in Line:32-33 of Fig. \ref{fig.5}. In this step, OFMaps move in $Y$ from PE partitions to the drain buffer.
Fig. \ref{fig.6}(b) describes the symbolic representation, and Fig. \ref{fig.6}(c) shows the loop-nest definition of the spatial column and temporal row assignment of OFMap data from two PE partitions to the drain buffer. In step \ding{184}, the $load$ input of all PEs is set to 0 and, $Mul\_En$ is set to 1, which indicates PEs operate in the calculate mode.

\begin{figure}[]
\centering
\centerline{\includegraphics[scale=0.85]{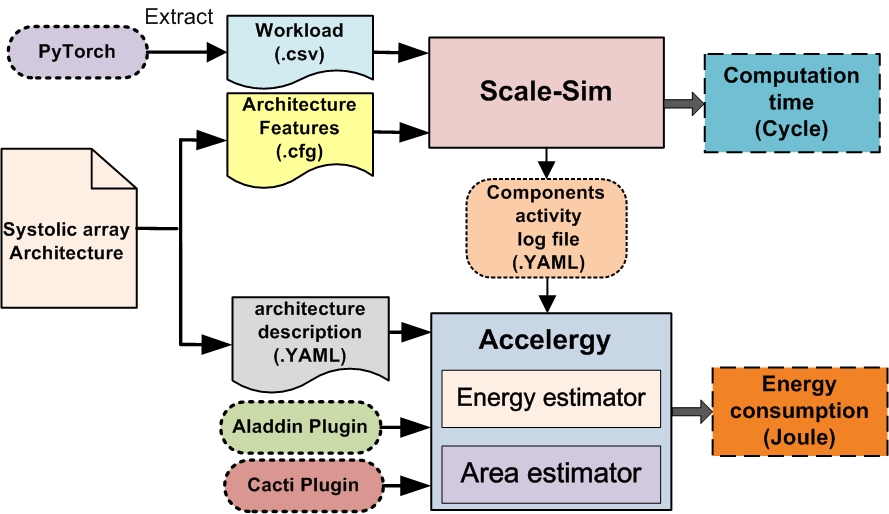}}
\caption{The Simulation toolchain}
\label{fig.8}
\end{figure}  
\begin{figure*}[] 
    \centering
    \includegraphics[width=\textwidth]{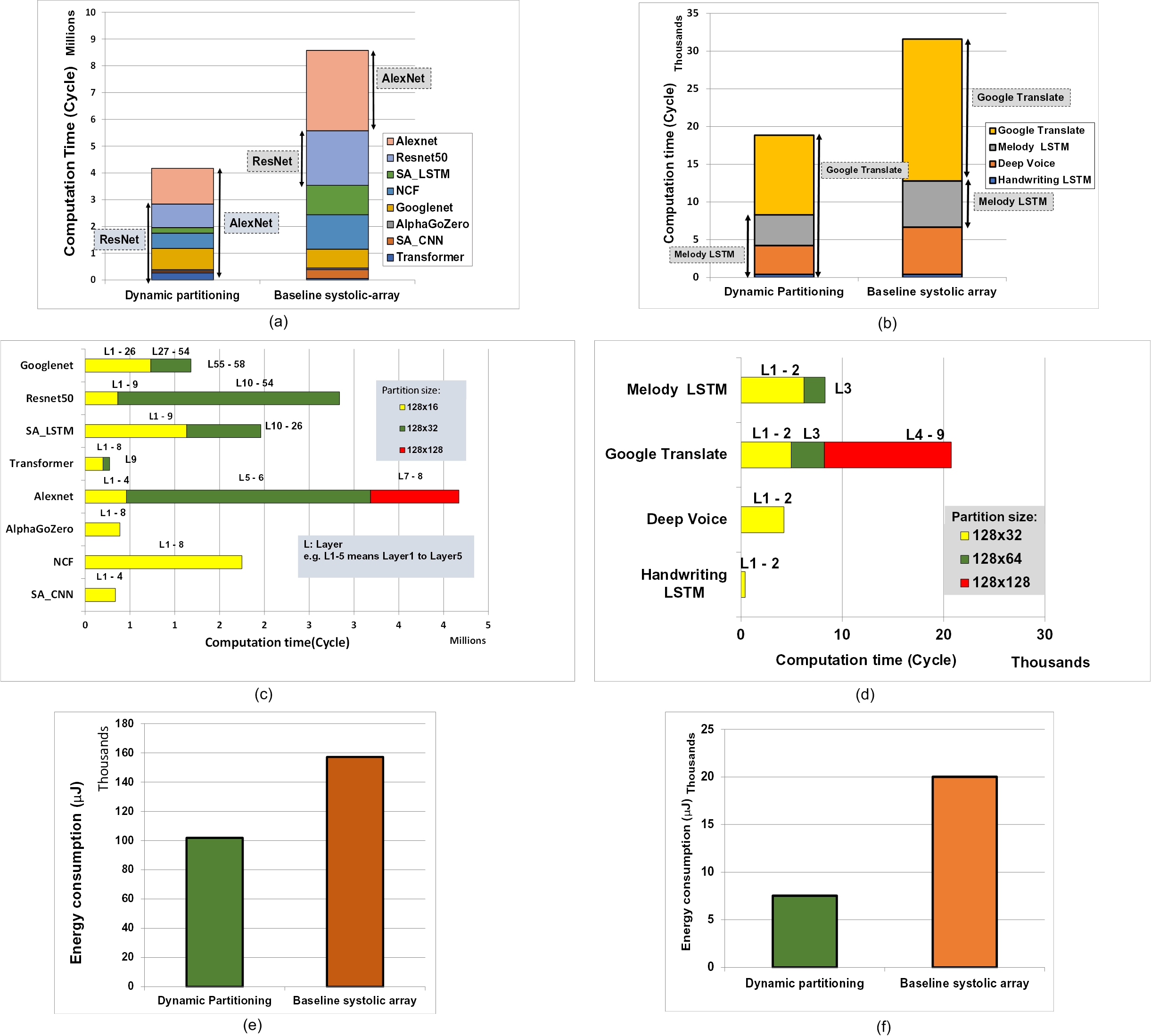}
    \caption{(a) The computation time of multi-domain workload (b) The computation time of RNN workload (c) The detailed computation time with assigned partition sizes of multi-domain workload (d) The detailed computation time with assigned partition sizes of RNN workload (e) The energy consumption of multi-domain workload (f) The energy consumption of RNN workload}
    \label{fig.9}
\end{figure*} 

\section{Simulation Results}
\subsection{Simulation Workload}
{\hskip 1em}We used 12 PyTorch DNN models \cite{pytorch} and divided them into two groups to simulate the multi-tenancy environment in \textit{heavy} and \textit{light} loads. Table 1, illustrates the type and load of the workloads. The first group of DNNs is the heavy-load multi-domain type, and the second is the light-load recurrent neural network (RNN) type. 

\subsection{Simulation Toolchain}
{\hskip 1em}We simulate a systolic array similar to TPU v3 which comprises $128\times128$ PEs with load, feed, and drain buffers. We used the \textit{Scale-Sim} simulator \cite{samajdar2020systematic} to evaluate the computation time, and the energy consumption is estimated based on  45nm technology by \textit{Accelergy} framework \cite{wu2019accelergy}. Scale-Sim is a dedicated systolic array simulator, and Accelergy is an open source tool for estimating accelerator area and energy consumption. As shown in Fig. \ref{fig.8}, we employ both tools by extracting component activities in the form of a logfile from the Scale-Sim and then importing the log file into the Accelergy framework as a component activity. Accelergy uses \textit{Cacti} \cite{li2011cacti} and \textit{Aladdin} \cite{shao2014aladdin} plugins to estimate energy usage based on component activity.

\begin{table}
\centering
\caption{Simulation workloads}
\scalebox{0.8}{
\begin{tabular}{c|c|c}
\textbf{Type}                                                                                             & \textbf{Domain}                       & \textbf{Training model}                                                           \\ 
\hline
\multirow{8}{*}{\begin{tabular}[c]{@{}c@{}}Multi-domain\\~/\\~Heavy load workload\end{tabular}}           & \multirow{3}{*}{Image classification} & AlexNet\cite{krizhevsky2017imagenet}                                                                       \\ 
\cline{3-3}
                                                                                                          &                                       & ResNet50\cite{he2016deep}                                                                      \\ 
\cline{3-3}
                                                                                                          &                                       & GoogleNet\cite{szegedy2015going}                                                                     \\ 
\cline{2-3}
                                                                                                          & \multirow{2}{*}{Sentiment analysis}   & \begin{tabular}[c]{@{}c@{}}Sentiment analysis CNN\\(SA\_CNN)\cite{santos2017sentiment}\end{tabular}     \\ 
\cline{3-3}
                                                                                                          &                                       & \begin{tabular}[c]{@{}c@{}}Sentiment analysis LSTM\\~(SA\_LSTM)\cite{wang2016dimensional}\end{tabular}  \\ 
\cline{2-3}
                                                                                                          & Recommendation system                 & \begin{tabular}[c]{@{}c@{}}Neural collaborative filter\\~(NCF)\cite{chen2019joint}\end{tabular}   \\ 
\cline{2-3}
                                                                                                          & Intelligent search                    & AlphGoZero\cite{silver2017mastering}                                                                    \\ 
\cline{2-3}
                                                                                                          & Natural language processing           & Transformer \cite{vaswani2017attention}                                                                  \\ 
\hline
\multirow{4}{*}{\begin{tabular}[c]{@{}c@{}}Recurrent Neural Network\\~/\\Lightload workload\end{tabular}} & Melody extraction                     & Melody LSTM \cite{park2017melody}                                                                  \\ 
\cline{2-3}
                                                                                                          & Language translation                  & Google Translate \cite{wu2016google}                                                             \\ 
\cline{2-3}
                                                                                                          & Text to speech                        & Deep voice\cite{arik2017deep}                                                                    \\ 
\cline{2-3}
                                                                                                          & Handwriting recognition               & Handwriting LSTM \cite{carbune2020fast}                                                            
\end{tabular}}
\end{table} 

\subsection{Results}
{\hskip 1em}Fig. \ref{fig.9}(a) and \ref{fig.9}(b) show the computation time of our two workloads in the baseline systolic array with no partitioning algorithm and a systolic array with the dynamic partitioning algorithm. As we can see in Fig. \ref{fig.9}(a) and \ref{fig.9}(b), all DNNs run sequentially in baseline scenario but concurrently in the dynamic partitioning algorithm.
DNNs such as AlexNet in multi-domain workloads or Google Translate in RNN workloads are completed later than other DNNs. However, these DNNs have been running in parallel with the other DNNs from the beginning, but the processing of DNNs with smaller dimensions is completed earlier.

Employing the dynamic allocation algorithm increases resource utilization, which reduces the total computation time due to the parallel execution of the layers. Fig. \ref{fig.9}(c) and \ref{fig.9}(d) 
show more details about assigned resource partitions to DNN layers.

According to Fig. \ref{fig.9}(c), the processing of AlphaGoZero, NCF, and SA\_CNN were completed using $128\times16$ partitions due to their low dimensional layers. The reason that all layers of NCF are processed by $128\times16$ partition, while $128\times16$ and $128\times32$ partitions process some layers of SA\_LSTM or Transformer, is that the NCF layers are lighter (have lower dimensions) compared with SA\_LSTM or Transformer DNNs.

As per our algorithm in Fig. \ref{fig.5}, the allocated partitions become free once each layer is finished and can be merged with other adjacent free partitions to create a partition with more resources to speed up the remaining DNNs.

More complex DNNs with high dimension layers like ResNet50, GoogleNet, SA\_LSTM, and Transformer use both $128\times16$ and $128\times32$ partitions, and the last two fully connected layers of AlexNet use all PEs to complete the processing. In the RNN workload (Fig. \ref{fig.9}(d)), we see a similar case in which the handwriting LSTM and Deep voice DNNs are processed by a $128\times32$ partition. The last layer of Melody LSTM uses a $128\times64$ partition size, and the last six layers of Google translate use all PEs to finish the processing.

Fig. \ref{fig.9}(e) and \ref{fig.9}(f) show energy consumption of the baseline systolic array and the dynamic partitioning technique respectively. As shown in Fig. \ref{fig.9}(e) and Fig. \ref{fig.9}(f), utilizing a dynamic partitioning mechanism saves 35\% in multi-domain workload and 62\% in RNN workload compared to the baseline systolic array. Energy saving is achieved due to the efficient resource utilization of our dynamic partitioning algorithm.

\section{Related Work}


{\hskip 1em}Multi-tenancy has been addressed for DNN accelerators in server and edge platforms. There are two main types of hardware accelerators for DNNs. \textit{Spatial accelerators} consist of a network of processing elements connected by some sort of network on chip (NoC). Spatial accelerators provide a lot of flexibility in the division of work between PEs. In contrast, systolic arrays provide a fixed pattern of execution and communication between PEs. Systolic arrays can be extremely efficient because the PEs and intercommunication network are extremely simple and deterministic. In contrast, spatial accelerators provide much greater flexibility, at the cost of more complex hardware to support the flexibility.

Spatial accelerators are arguably much better suited to multi-tenancy in edge devices because they provide the flexibility for groups of PEs to work independently of others. Supporting multi-tenancy on a systolic array is much more difficult because the PEs work together in lock-step with a fixed data movement pattern. Where systolic array-based accelerators support multi-tenancy, it is almost always by providing separate systolic arrays for each tenant, rather than splitting one systolic array between tenants. In contrast, our work focuses on dividing a single systolic array between multiple tenants.

\textbf{Edge-side multi-DNN processing by spatial accelerators.} In edge platforms, DNN types and numbers are commonly fixed and predefined. Hence, a designer can modify an accelerator design according to the type of DNNs. For example, Herald \cite{kwon2021heterogeneous} consists of multiple accelerators called sub-accelerators tuned for different dataflows to support multi-DNN processing. The sub-accelerators are divided among the tenants so that there is no need to share sub-accelerators.

\textbf{Server-side multi-tenancy by systolic array based accelerators.} Planaria \cite{ghodrati2020planaria} and AI-MT \cite{baek2020multi} are examples of server-side acceleration. Planaria comprises sixteen systolic
arrays that are connected via an on-chip bi-directional ring bus. Planaria uses an extension of the systolic array called \textit{omni-directional}. In the omni-directional type, the PE logic is modified to send the calculated partial sum to the output ports in all directions. Thus, the modified PE logic routes the tensor values in four directions, which turns the systolic array into a more general full spatial accelerator. Lee et al. proposed dataflow mirroring \cite{lee2021dataflow} for an omni-directional systolic array, to compute multiple DNN layers by different inter-PE communication patterns. 
As previously mentioned, we partition a single systolic array, and we require only a very small additional hardware cost compared with solutions like utilizing an omni-directional structure.

AI-MT enhanced the AI-multitasking hardware to a systolic array-based accelerator to divide each layer into sub-layers and categorize them into compute or memory-intensive types. Subsequently, the scheduler schedules sub-layers based on their type and runs them sequentially. In other words, AI-MT performs task scheduling at the hardware level.

Google performs multi-tenant inference and training operations on a supercomputer of connected Tensor Processing Units (TPU) \cite{jouppi2020domain}.
Each TPU is a systolic array, but there is no support for partitioning the TPU. So, multi tenancy is performed by allocating different tenant DNNs to different TPUs.
TPU v2 and v3 \cite{norrie2021design} comprise pods where each pod consists of eight TPU boards, and each TPU board includes eight systolic arrays. Thus, a TPU board is the primary cluster of systolic chips. 



\section{Conclusion}
{\hskip 1em}Multi-tenancy and multi-DNN processing are the new challenges in cloud and edge devices that affect computation time and energy consumption. This paper proposes a dynamic resource partitioning algorithm with a slight hardware modification and uses a data-level method to share an accelerator across several DNNs. 
Therefore, multiple DNNs use partitions of memory and processing components as sub-accelerators.
We divide the PEs of the weight-stationary systolic array into vertical partitions that achieve data reuse in multi-tenant processing. At the hardware level we add a tri-state gate to the PE logic to allow data to flow through PEs without triggering computation.
We apply two different types and loads of multi-DNN workloads to simulate the proposed idea. Simulation results show significant improvements in computation time and energy consumption due to increased resource utilization.

\section{Acknowledgment}
{\hskip 1em}This work was supported, in part, by Science Foundation Ireland under grant No. 13/RC/2094\_P2, co-funded under the European Regional Development Fund through the Southern \& Eastern Regional Operational Programme to Lero \footnote{\href{https://lero.ie/}{https://lero.ie/}} (Science Foundation Ireland Research Centre for Software), and, in part, this project has received funding from the European Union’s Horizon 2020 research and innovation programme under the Marie Skłodowska-Curie grant agreement No 754489.

\bibliographystyle{unsrtnat}
\bibliography{references}  
\end{document}